\documentclass[aps,prd,amssymb,eqsecnum,showpacs,showkeyes,secnumarabic,graphics,floatfix,nofootinbib,tightenlines,longbibliography]{revtex4}
\usepackage{graphicx}
\usepackage{bm}
\usepackage{cancel}
\usepackage{epsfig}
\usepackage{dcolumn}
\usepackage{amsmath}
\usepackage{hhline}
\usepackage[utf8]{inputenc}
\usepackage[dvipsnames]{xcolor}
\usepackage[breaklinks=true,colorlinks=true,
linkcolor=blue,urlcolor=Blue,citecolor=MidnightBlue,
bookmarks=true,bookmarksopenlevel=2]{hyperref}
\usepackage{bm}
\usepackage{caption}
\usepackage{amsmath}
\usepackage {amssymb}

\def \beq  {\begin{equation}}
\def \eeq  {\end{equation}}
\def \ber  {\begin{eqnarray}}
\def \eer  {\end{eqnarray}}
\def \Geff {G_{\rm eff} }
\def \Geffz {$G_{\rm eff}(z)$ }

\def \Geff {G_{\rm eff}}

\begin{document}
\newcommand{\newc}{\newcommand}

\newc{\be}{\begin{equation}}
\newc{\ee}{\end{equation}}
\newc{\ba}{\begin{eqnarray}}
\newc{\ea}{\end{eqnarray}}
\newc{\bea}{\begin{eqnarray*}}
\newc{\eea}{\end{eqnarray*}}
\newc{\D}{\partial}
\newc{\ie}{{\it i.e.} }
\newc{\eg}{{\it e.g.} }
\newc{\etc}{{\it etc.} }
\newc{\etal}{{\it et al.}}
\newc{\lcdm}{$\Lambda$CDM }
\newc{\lcdmnospace}{$\Lambda$CDM}
\newc{\wcdm}{wCDM }
\newc{\omom}{$\Omega_{0m}$ }
\newc{\omomnospace}{$\Omega_{0m}$}
\newc{\plcdm}{Planck18/$\Lambda$CDM }
\newc{\plcdmnospace}{Planck18/$\Lambda$CDM}
\newc{\wlcdm}{WMAP7/$\Lambda$CDM }
\newc{\fs}{{\rm{\it f\sigma}}_8}
\newc{\fsz}{{\rm{\it f\sigma}}_8(z)}

\newcommand{\nn}{\nonumber}
\newc{\ra}{\Rightarrow}
\title{Is gravity getting weaker at low z?\\
Observational evidence and theoretical implications}

\author{Lavrentios Kazantzidis}\email{lkazantzi@cc.uoi.gr}
\affiliation{Department of Physics, University of Ioannina, GR-45110, Ioannina, Greece}
\author{Leandros Perivolaropoulos}\email{leandros@uoi.gr}
\affiliation{Department of Physics, University of Ioannina, GR-45110, Ioannina, Greece}

\date{\today}

\begin{abstract}
Dynamical observational probes of the growth of density perturbations indicate that gravity may be getting weaker at low redshifts $z$. This evidence is at about $2-3\sigma$ level and comes mainly from weak lensing data that measure the parameter $S_8=\sigma_8 \sqrt{\Omega_{0m}/0.3}$  and redshift space distortion data that measure the growth rate times  the amplitude of the linear power spectrum parameter $\fsz$.  The measured $\fs$ appears to be lower than the prediction of General Relativity (GR) in the context of the standard  \lcdm model as defined by the Planck best fit parameter values. This is the well known $\fs$ tension of \lcdmnospace, which constitutes one of the two main large scale challenges of the model along with the $H_0$ tension.  We review the observational evidence that leads to the $\fs$ tension and discuss some theoretical implications. If this tension is not a systematic effect it may be an early hint of modified gravity with an evolving effective Newton's constant $G_{eff}$ and gravitational slip parameter $\eta$. We discuss such best fit parametrizations of $\Geff(z)$ and point out that they can not be reproduced by simple scalar-tensor and $f(R)$ modified gravity theories because these theories generically predict stronger gravity  than General Relativity (GR) at low $z$ in the context of a \lcdm background $H(z)$. Finally, we show weak evidence for an evolving reduced absolute magnitude of the SnIa of the Pantheon dataset at low redshifts ($z<0.1$) which may also be explained by a reduced strength of gravity and may help resolve the $H_0$ tension.
\end{abstract}
\maketitle

\section{Introduction}
\label{sec:Introduction}

The simplest model consistent with current cosmological observations
is the \lcdm model, which assumes the existence of a fine tuned cosmological constant that drives the accelerating expansion of the Universe \cite{Carroll:2000fy}.  A wide range of cosmological observations have imposed strong constraints on the six free parameters of the model. These observations include Type Ia supernovae (SnIa) used as distance indicators \cite{Riess:1998cb,Perlmutter:1998np,Betoule:2014frx,Scolnic:2017caz}, the Cosmic Microwave Background (CMB) angular power spectrum \cite{Hinshaw:2012aka,Ade:2015xua,Aghanim:2018eyx}, the Baryon Acoustic Oscillations (BAO) \cite{Aubourg:2014yra,Alam:2016hwk}, Cluster Counts (CC) \cite{Rozo:2009jj,Rapetti:2008rm,Ade:2015fva,Bocquet:2014lmj,Ruiz:2014hma}, Weak Lensing (WL) \cite{Hildebrandt:2016iqg,Joudaki:2017zdt,Troxel:2017xyo,Kohlinger:2017sxk,Abbott:2017wau,Abbott:2018xao} and Redshift Space Distortions (RSD) \cite{Macaulay:2013swa,Johnson:2015aaa,Basilakos:2016nyg,Nesseris:2017vor,Kazantzidis:2018rnb}.
The first three of the above (SnIa, CMB power spectrum peak locations and BAO), act as cosmological distance indicators and directly probe the cosmic metric independent of the underlying theory of gravity. These are known as ``geometric probes" \cite{Nesseris:2006er,Basilakos:2013nfa,Ruiz:2014hma}. The other three types of observations, probe simultaneously the cosmic metric and the growth rate of cosmological perturbations. They are  sensitive to the dynamics of growth and thus to the type of the underlying theory of gravity. These are known as ``dynamical probes" \cite{Nesseris:2006er,Basilakos:2013nfa,Ruiz:2014hma}.

\begin{figure}[!h]
\centering
\includegraphics[width = 0.5\textwidth]{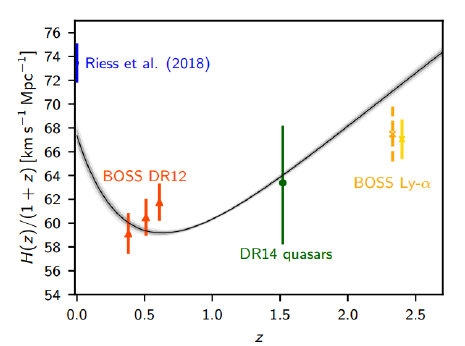}
\caption{The comoving Hubble parameter as a function of $z$ superimposed with BAO data from BOSS DR12 \cite{Alam:2016hwk} survey (orange points), BOSS DR14 quasar sample \cite{Zarrouk:2018vwy} (green point), SDSS DR12 Ly$\alpha$ sample \cite{Bautista:2017zgn}  (yellow points) and the Hubble Space Telescope survey \cite{Riess:2018uxu} (blue point). The black line corresponds to the best fit obtained from the Planck18 CMB data under the assumption of a \lcdm background, while the grey areas are the $1 \sigma$ regions (from Ref. \cite{Aghanim:2018eyx}).}
\label{fig:Hubblez}
\end{figure}

The consistency of the standard \lcdm model with cosmological observations requires that the model passes two types of tests:
\begin{itemize}
\item
The quality of fit of the model is acceptable in the context of each one of the above observational probes.
\item
The best fit values of the six free parameters of the model obtained with each individual probe are consistent with each other at a level of about $1-2\sigma$. If this is not the case, then we have a ``tension" of \lcdmnospace.
\end{itemize}
The \lcdm model appears to pass the first test in the context of practically all current observational probes. However, there seem to be some issues for \lcdm in the context of the second test \cite{Bull:2015stt}. In particular, two classes of tension have appeared to persist and amplify during the past decade. The first is the $H_0$ tension, where $H_0$ is the Hubble parameter.  The Planck mission \cite{Ade:2015xua,Aghanim:2018eyx} reports that $H_0=67.4 \pm 0.5 \; km \, s^{-1} \, Mpc^{-1}$ at the $1 \sigma$ level, whereas local measurements mainly from  Cepheid \cite{Tammann:2008xf} and SnIa luminosity distance indicators \cite{Riess:2019cxk,Riess:2016jrr} report that $H_0=74.03 \pm 1.42 \; km \, s^{-1} \, Mpc^{-1}$ at the $1 \sigma$ level, a value approximately $4\sigma$ away from the Planck reported one. This tension indicates that the local measurement of the Hubble parameter (at scales up to $400 Mpc$) obtained mostly using SnIa, is higher than the global value obtained from the Hubble volume on scales of $10 \, Gpc$ \cite{Margalef-Bentabol:2012kwa} through an extrapolation of $H(z)$ from the last scattering surface to the present time in the context of \lcdm scenario, as shown in Fig. \ref{fig:Hubblez}. 

Possible explanations of this tension include systematic errors of the CMB and/or the SnIa distance indicators. Alternatively this tension could be an early hint for physical deviations from the \lcdm model (see e.g. Ref. \cite{Huterer:2017buf} for a recent review). The later possibility is more likely in view of the fact that other local cosmological observations including other SnIa data analysis methods \cite{Efstathiou:2013via,Cardona:2016ems,Zhang:2017aqn},  gravitational lensing \cite{Suyu:2012aa} and Tully-Fisher type calibration of SnIa \cite{Sorce:2012pk}, appear to be consistent with the SnIa measurement \cite{Riess:2019cxk,Riess:2016jrr}. In contrast, measurements involving BAO \cite{Aubourg:2014yra} and SnIa calibration using the tip of the red-giant branch distances \cite{Tammann:2012ut}  are consistent with the extrapolated global CMB measurements of $H_0$. 

A natural cause  for this tension could be cosmic variance. If it happens that we live in a locally underdense region of the Universe we would locally measure a value of $H_0$ that would be higher than the mean value over the whole Universe. It has been shown however \cite{Wu:2017fpr,Kazantzidis:2020tko}, that the required magnitude of such a underdensity  on the required scales of $150  \, Mpc$ is very unlikely in a \lcdm universe. In such a Universe cosmic variance adds a $1\sigma$ error to the locally measured $H_0$ of only $\sigma_{H_0}=0.31 \; km \, s^{-1} \, Mpc^{-1}$, which is negligible compared to the $6 \; km \, s^{-1} \, Mpc^{-1}$ needed to resolve the $H_0$ tension.

Non-gravitational physical mechanisms that can reduce the $H_0$ tension include the following:
\begin{itemize}
\item
Modifications of expansion rate at late times in the context of alternative dark energy models \cite{Yang:2018qmz,Yang:2019jwn,Belgacem:2017cqo,Alestas:2020mvb}, decaying dark matter models \cite{Pandey:2019plg,Vattis:2019efj}, or the presence of massive sterile neutrinos \cite{Zhao:2017urm} that tend to amplify the accelerating expansion at late times. Such modifications could drive upward the low $z$ part of the $H(z)$ curve shown in Fig. \ref{fig:Hubblez}, thus bringing the $z=0$ prediction of the CMB closer to the $H_0$ result of the local measurements of Ref. \cite{Riess:2016jrr}
\item
Inhomogeneous cosmologies \cite{Lukovic:2018ljo} that would make a local deep underdensity more likely than in the case of \lcdmnospace.
\item 
A new component of dark radiation \cite{Bernal:2016gxb} that would tend to decrease the sound horizon $r_s$ at radiation drag, thus leading to a predicted increase of $H_0$ by shifting the whole curve of Fig. \ref{fig:Hubblez} upwards. This approach has the advantage of shifting at the same time the BAO points shown in Fig. \ref{fig:Hubblez}. 
\end{itemize}
The origin of these mechanisms is non-gravitational and the consensus is that since $H_0$ is a geometric parameter it can not be affected by modifications of GR. However, as discussed in more detail below, the physics of SnIa is heavily based on the assumption of validity of GR. For example, an evolving Newton's constant at low redshifts would directly affect the absolute magnitude of SnIa, leading to requirement for a new interpretation of the SnIa distance moduli. Therefore, even though the SnIa absolute magnitude is usually assumed constant and is marginalized as being a nuisance parameter, its possible evolution may carry useful information about the robustness of the determination of $H_0$ using SnIa and about possible modifications of GR. Two interesting questions therefore arise: 
\begin{itemize}
\item
Are there indications for evolution of the SnIa absolute magnitude at low $z$?
\item
What would be the implications of such evolution on the derived value of $H_0$ and on the possible evolution of the effective Newton's constant?
\end{itemize}
These are among the questions discussed in what follows. 

The second tension in the context of \lcdm is the $\sigma_8$ tension, where $\sigma_8$ is the density rms matter fluctuations within spheres of radius $8 h^{-1} Mpc$ and is determined by the amplitude of the primordial fluctuations power spectrum and by the growth rate of cosmological fluctuations. In particular, dynamical probes, (mainly RSD \cite{Macaulay:2013swa,Nesseris:2017vor,Basilakos:2017rgc,Kazantzidis:2018rnb}, WL data \cite{Hildebrandt:2016iqg,Kohlinger:2017sxk,Joudaki:2017zdt,Abbott:2017wau} and $E_G$ data \cite{Skara:2019usd}), favor lower values of $\sigma_8$ and/or \omom than the corresponding values reported by Planck \cite{Ade:2015xua,Aghanim:2018eyx} at a $2-3\sigma$ level. This tension, if not due to systematics of the dynamical probes or CMB data, could be interpreted as an indication for a weaker gravitational growth of perturbations than the growth indicated by GR in the context of a \lcdm model with the \plcdm parameter values, which are shown in the following Table \ref{PKtab:plcdm18}

\begin{table}[h!]
\caption{\plcdm  parameters values from Ref. \cite{Aghanim:2018eyx} based on TT,TE,EE, lowE and lensing likelihoods.}
\label{PKtab:plcdm18}
\begin{centering}
\begin{tabular}{cc}
 \hline 
 \rule{0pt}{3ex}  
  Parameter & \plcdm \cite{Aghanim:2018eyx}\\
    \hline
    \rule{0pt}{3ex}  
$\Omega_b h^2$ & $0.02237 \pm 0.00015$ \\
$\Omega_c h^2$ & $0.1200 \pm 0.0012$  \\
$n_s$ & $ 0.9649 \pm 0.0042$\\
$H_0$ & $67.36 \pm 0.54$ \\
$\Omega_{0m}$ & $0.3153 \pm 0.0073$ \\
$w$ & $-1$  \\
$\sigma_8$ & $ 0.8111 \pm 0.0060$\\
\hline
\end{tabular}
\end{centering}
\end{table}

In addition to a possible evolution of the effective Newton constant discussed below, non-gravitational mechanisms can also reduce the $\sigma_8$ tension (see \eg Ref. \cite{Ishak:2018his} for a  recent review). Such effects include the following:
\begin{itemize}
\item
Interacting dark energy models, which modify the equation for the evolution of linear matter fluctuations in a given $H(z)$ cosmological background \cite{Pourtsidou:2016ico,Barros:2018efl,Camera:2019vbp}.
\item
Dynamical dark energy models \cite{Melia:2016djn,Lambiase:2018ows,Ooba:2018dzf,Yang:2018qmz,Joudaki:2016kym,Barros:2018efl} and running vacuum models  \cite{Gomez-Valent:2017idt,Gomez-Valent:2018nib},  which modify the cosmological background $H(z)$ to a form different from \lcdmnospace.
\item
Effects of massive neutrinos \cite{Joudaki:2016kym,DiazRivero:2019ukx}.  Neutrinos are relativistic at early times (contribute to radiation) while at late times they become non-relativistic but with significant velocities (hot dark matter) while they constitute a non-negligible fraction of the dark matter of the universe. The conversion of radiation
to hot dark matter plays a role in the Hubble expansion.  At the same time the residual streaming velocities are still large enough
at late times to slow down the growth of structure.
Thus, neutrinos affect both background expansion and the growth of cosmological perturbations in such a way as to slow down the growth as required by the RSD data. Their effects on easing the $\sigma_8$ tension coming from WL data has been questioned by the recent analysis of Ref. \cite{DiValentino:2018gcu} (see also Fig. \ref{fig:S8const}) 

\begin{figure}[!h]
\centering
\includegraphics[width = 0.8\textwidth]{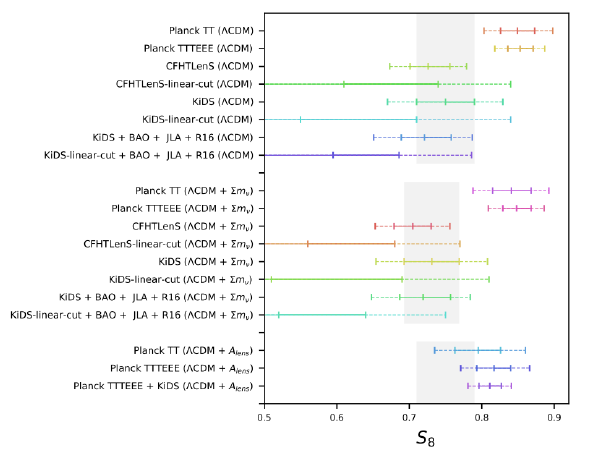}
\caption{$1-2 \sigma$ constraints for $S_8=\sigma_8 \sqrt{\frac{\Omega_{0m}}{0.3}}$ for various combinations of datasets and models superimposed with the Kilo Degree Survey (KiDS) \cite{Joudaki:2016kym} survey bounds (grey regions) for each cosmological model (adopted from Ref. \cite{DiValentino:2018gcu}). In particular, the CFHTLenS linear cut model corresponds to the conservative cut of the cosmic shear data of the Canada-France-Hawaii Telescope Lensing (CFHTLens) survey \cite{Heymans:2012gg,Erben:2012zw} in order to reduce the non-linear scale contribution \cite{Ade:2015rim}, the JLA acronym corresponds to the Supernovae Data from Ref. \cite{Betoule:2014frx}, the R16 stands for the $H_0$ measurement given in Ref. \cite{Riess:2016jrr}, the BAO data correspond to the data used in Ref. \cite{Ade:2015xua}, while the KiDS linear cut model describes the conservative cut of Ref. \cite{Joudaki:2017zdt} and $\sum m_\nu$ stands for the inclusion of massive neutrinos. Notice that only the introduction of the $A_{lens}$ parameter which is degenerate with the evolution of $\Geff$ \cite{DiValentino:2015bja} can lead to a reduction of the tension between Planck and dynamical probes.}
\label{fig:S8const}
\end{figure} 
\end{itemize}
Besides these categories, alternative parameters beyond the standard ones, such as a running scalar spectral index, a modified matter expansion rate or a bulk viscosity coefficient may have the potential to ease the $\sigma_8$ tension \cite{Wang:2019acf}. 

In addition to these non-gravitational effects that can slow down growth at low redshifts, modified gravity theories can also contribute in the same direction in a manner that is more generic and fundamental. The feature required from this class of theories is a reduced effective Newton's constant $\Geff$ at low redshifts. It turns out that this behavior can not be achieved in a \lcdm background for most scalar-tensor and $f(R)$ theories \cite{Tsujikawa:2007gd,Gannouji:2018ncm,Nesseris:2017vor}. However, it is possible in other less generic modified gravity theories including telleparallel theories of gravity \cite{DAgostino:2018ngy,Gonzalez-Espinoza:2018gyl}, Horndeski theories \cite{Kennedy:2018gtx,Linder:2018jil,Gannouji:2020ylf} or theories beyond Horndeski  \cite{DAmico:2016ntq}.

Clearly a reduced (compared to GR)  evolving effective Newton's constant would have important signatures on low $z$ cosmological observations. In particular:
\begin{itemize}
\item
It would affect \cite{Nesseris:2017vor,Kazantzidis:2018rnb} the low $l$ CMB power spectrum through the Integrated Sachs-Wolfe (ISW) effect \cite{Pogosian:2005ez,Ho:2008bz}.
\item
It would affect \cite{Nesseris:2017vor,Kazantzidis:2018rnb} the growth rate of cosmological fluctuations as detected through the RSD \cite{Macaulay:2013swa,Johnson:2015aaa,Basilakos:2016nyg,Nesseris:2017vor,Kazantzidis:2018rnb}, WL \cite{Hildebrandt:2016iqg,Joudaki:2017zdt,Troxel:2017xyo,Kohlinger:2017sxk,Abbott:2017wau,Abbott:2018xao} and CC data \cite{Rozo:2009jj,Rapetti:2008rm,Ade:2015fva,Bocquet:2014lmj,Ruiz:2014hma}.
\item
It would induce an evolution of the SnIa absolute magnitude which depends on the magnitude of Newton's constant \cite{Amendola:1999vu,Gaztanaga:2001fh,Nesseris:2006jc,Wright:2017rsu,Sapone:2020wwz}. Notice, however, that the value of the effective Newton's constant here should be obtained from a strong gravity calculation in the context of a modified gravity theory and thus is not in general identical to the $\Geff$ derived for the growth of cosmological perturbation that involves a perturbative calculation.
\end{itemize}
For a viable modified gravity mechanism there should be consistency with respect to the type and magnitude of Newton's constant evolution favored by the above cosmological observations, keeping in mind the strong gravity effects involved in the SnIa physics. It will be seen in what follows that indeed all of the above probes mildly favor a reduced value of Newton's constant at low $z$. However, the favored magnitude and statistical significance of such a reduction varies  among the above observational probes.

A Newton's constant \Geffz evolving with redshift, may be parametrized in the context of a wide range of parametrizations including theoretically motivated \cite{Bertschinger:2008zb,DiValentino:2015bja,Ade:2015rim,Baker:2014zva,Li:2018tfg} and model independent \cite{Nesseris:2017vor,Kazantzidis:2018rnb} forms and leads to a modification of the linear growth of cosmological perturbations. 

This modified growth equation is obtained by considering the perturbed Friedmann–Lema\^itre-Robertson–Walker (FLRW) metric in the Newtonian gauge which is given by \cite{Bardeen:1980kt,Ma:1995ey,EspositoFarese:2000ij}
\be
ds^2 = -(1+2\Psi)dt^2 + a^2(1-2\Phi) d{\vec{x}}^2\ , \label{PKeq:pertfrw}
\ee
where $a$ is the scale factor that is connected to the redshift $z$ through $a=1/(1+z)$ and $\Psi$, $\Phi$ correspond to the Bardeen potentials in the Newtonian gauge \cite{Ma:1995ey}. Einstein's equations in  Fourier space at linear order take the form \cite{Huterer:2013xky,Pogosian:2010tj,Perenon:2019dpc}
\ba
 k^2 \Psi &=& -4\pi G_N \mu (a, k) a^2 \rho \Delta \ ,
\label{PKeq:poissonmg1} \\
 \Phi &=& \eta(a, k) \Psi \ ,
\label{PKeq:poissonmg2} \\
 k^2(\Phi +\Psi) &=& -8\pi G_N\, \Sigma(a, k)\, a^2 \rho \Delta \ ,
\label{PKeq:poissonmg3}
\ea
where $\mu\equiv \Geff/G_{\textrm{N}}$ ($G_N$ is Newton's constant as measured by local experiments and Solar System observations), $\Delta$ is the comoving density contrast defined as $\Delta \equiv \delta+3 a H u/k$. $\delta \equiv \frac{\delta \rho}{\rho}$ is the linear matter growth  factor, $u$ is the irrotational component potential of the peculiar velocity, $\rho$ is the  matter density of the background, $\eta$ is the gravitational slip and $\Sigma\equiv G_{\textrm{L}}/G_{\textrm{N}}$ is the lensing normalized Newton constant.

In GR, $\Geff$, which is connected with the growth of matter  perturbations and $G_L$, which is related with the lensing of light through the Weyl potential $\frac{\Phi_+}{2}=\Phi+\Psi$, coincide with $G_N$. The Weyl potential can be connected with lensing since  the cosmic convergence of null geodesics with respect to unperturbed geodesics is given by \cite{Tereno:2010dt}
\be
\kappa(r,\theta)=\frac{1}{4}\int_0^{r} dr'\,\frac{r-r'}{r}r'
\nabla^2\Phi_+(\theta,r')
\ee
where $r$ and $\theta$ are the comoving coordinates of the source.
The parameters $\mu$, $\eta$ and $\Sigma$ are connected as 
\be 
\Sigma(a, k)=\frac{\mu(a, k)\, \left[1+\eta (a, k) \right]}{2} \ ,
\ee
and they are key parameters in detecting deviations from GR where their value coincides with unity. 

In what follows we focus on the parameter $\mu$. This parameter is associated with the linear matter growth  factor through the growth equation \cite{Perenon:2019dpc}
\be 
\ddot{\delta} +2H\dot{\delta}= \frac{3}{2}\,H^2\,\Omega_m \,\mu \, \delta \ ,
\label{deltatevol}
\ee
where the dot denotes differentiation with respect to cosmic time $t$. Eq. (\ref{deltatevol}) is derived using the conservation of the matter energy momentum tensor and the modified Poisson eq. \eqref{PKeq:poissonmg1}, assuming scales much smaller than the Hubble scale.  For most modified gravity models the scale dependence of $\mu$ is very weak in scales much smaller than the Hubble scale. In addition, most growth data do not report scale dependence but only redshift dependence.  Thus, we only parametrize the dependence of  $\mu$ on the the scale factor, \ie $\mu(a,k)=\mu(a)$ and  we get the growth equation in redshift space as
\be
\delta'' +\left(\frac{H'}{H}-\frac{1}{1+z}\right)\, \delta'  =\frac{3}{2}\frac{\Omega_m }{\left(1+z\right)^2} \,\mu\, \delta \ , \label{PKeq:oddD}
\ee
where in eq. \eqref{PKeq:oddD}, the prime denotes differentiation with respect to the redshift $z$.  The equation for the growth rate $f(z)\equiv \frac{dln\delta}{dlna}$ may also be obtained from eq. \eqref{PKeq:oddD} as 
\be 
\left(1+z\right) f'-f^2+\left[ (1+z)\frac{H'}{H}-2 \right] f=-\frac{3}{2}\, \Omega_m \,\mu\ , \label{PKeq:oddf}
\ee
Fixing the background $H(z)$ and considering a specific parametrization for $\mu$, eq. \eqref{PKeq:oddf} can be solved (either numerically or analytically) with initial conditions deep in the matter era where $\delta\sim a$ (assuming GR is restored at early times). Combining this solution with the rms density fluctuations on scales of $8Mpc$, $\sigma_8$, which evolves as $\sigma_8(z)= \sigma_8 (z=0) \frac{\delta(z)}{\delta(z=0)}$, we obtain  theoretical prediction for the product $\fs$ given $\sigma_8(z=0)\equiv \sigma_8$, $H(z)$ and $\mu(z)$. In particular  
\be 
\fs(a)\equiv f(a)\cdot \sigma(a)=\frac{\sigma_8}{\delta(1)}~a~\delta'(a) \label{PK:eqfs8}
\ee 
is reported by many surveys since 2006, leading to collections of data which can be used to constrain simultaneously $H(z)$ and $\mu(z)$.

While $H(z)$ is usually parametrized as $wCDM$ \ie
\be
H^2(z)=H_0^2 \left[\Omega_{0m}(1+z)^3 +(1-\Omega_{0m})(1+z)^{3(1+w)}\right] \label{PKeq:wcdm}
\ee
which reduces to \lcdm for an  equation of state parameter $w=-1$, the effective Newton's constant parameter $\mu$ does not have a commonly accepted parametrization. Some authors motivated from the predictions of scalar-tensor theories use a scale dependent parametrization for $\mu$ and $\eta$ as \cite{Bertschinger:2008zb,Li:2018tfg}
\ba
\mu(a,k) &=& \frac{1+\beta_1 \lambda_1^2 k^2 a^s}{1+\lambda_1^2k^2a^s}, \nn \\
\eta(a,k) &=& \frac{1+\beta_2 \lambda_2^2 k^2 a^s}{1+\lambda_2^2k^2a^s} \label{PKeq:muetabzmod}
\ea

\noindent In the special case of $f(R)$ theories the parameters that appear in eq.  \eqref{PKeq:muetabzmod} are 
\ba 
\beta_1=4/3; \  \ \ \beta_2=1/2; \ \ \ \lambda_2^2/\lambda_1^2=4/3. 
\ea

\noindent Clearly for $\beta_1>1$ (as is the case for $f(R)$ theories \cite{Hu:2007nk,Starobinsky:2007hu} and in most scalar-tensor theories) we have $\mu>1$ at low $z$ and thus gravity is stronger than in GR at low $z$ in these classes of theories \cite{Hu:2007nk,Starobinsky:2007hu,Tsujikawa:2015mga,Gannouji:2018ncm,Polarski:2016ieb}.

Another parametrization that has been studied in the literature is the parametrization of no-slip gravity \cite{Linder:2018jil}, a subclass of Horndeski theories which remains viable after the binary star collision GW170817 \cite{TheLIGOScientific:2017qsa}. In this case $\mu$ takes the form \cite{Linder:2018jil}
\be 
\mu=\frac{2}{2+b+b \, tanh \left[\frac{\tau}{2} \, log_{10}(\frac{a}{a_t})\right]} \label{PK:eqnoslipmu}
\ee
where $b, \tau$ and $a_t$ correspond to parameters that describe  the amplitude, the rapidity and the scale factor at the time when $\mu$ shifts from unity in the early Universe to $\mu=1+b$.

An alternative scale dependent class of parametrizations for $\mu$ and $\eta$ is of the form \cite{Ade:2015rim,DiValentino:2015bja} 

\ba
\mu(a,k) &=& 1 + f_1(a) \frac{1 + c_1 (\lambda H/k)^2}{1+(\lambda H/k)^2}; \\
\eta(a,k) &=& 1 + f_2(a) \frac{1 + c_2 (\lambda H/k)^2}{1+(\lambda H/k)^2},
\ea

\noindent For sub-Hubble scales this parametrization becomes scale independent and has been expessed as \cite{DiValentino:2015bja}

\ba 
\mu(a,k) &=& 1 + E_{\rm{11}} \Omega_{\rm{DE}}(a)\,; \\
\eta(a,k) &=& 1 + E_{\rm{22}} \Omega_{\rm{DE}}(a)\,.
\ea
where $\Omega_{\rm{DE}}(a)$ is the density parameter of the dark energy. For $E_{11}<0$ gravity is weaker compared to GR at low $z$ and indeed the best fit value obtained in Ref. \cite{DiValentino:2015bja} when dynamical probes are taken into account is negative ($E_{11}=-0.21^{+0.19}_{-0.45}$ when CMB and WL data are taken into account).

A model and scale independent parametrization \cite{Nesseris:2017vor,Kazantzidis:2018rnb} for $\mu$ which reduces to the GR value at low and high $z$, while respecting the constraints from Solar Systems tests and from the nucleosynthesis \cite{Gannouji:2006jm,Nesseris:2006hp} is
\be
\mu =1+g_a(1-a)^n - g_a(1-a)^{2n}= 1+g_a \left(\frac{z}{1+z} \right)^n - g_a \left(\frac{z}{1+z} \right)^{2n}\ , \label{PKeq:geffansatz}
\ee
where $g_a$ and $n$ integer with $n\geq 2$ are parameters to be fit from data. A distinguishing feature of this parametrization is that it naturally and generically respects solar system and nucleosynthesis constraints ($\frac{d\mu}{dz}\vert_{z=0}=0, \mu(z =0)=1$, $\mu(z\rightarrow \infty)=1$) \cite{Muller:2005sr,Nesseris:2006hp,Gannouji:2006jm,Pitjeva:2013chs}.

An alternative approach for the parametrization of deviations from GR is based directly on the growth rate $f$ of density fluctuations. The growth rate $f$ is usually parametrized using the``growth index" $\gamma$ as 
\be 
f(z)=\frac{dln\delta}{dlna} \approx \Omega_m(z)^\gamma
\ee
where $\gamma$ in most dark energy models based on GR is $\gamma \approx 0.55$. For many modified gravity theories this quantity is not constant and is parametrised instead as a function of the redshift $z$ (see \eg Ref. \cite{Gannouji:2008wt} and for updated observational constraints of this parameter Refs. \cite{Shafieloo:2018gin,Gannouji:2018col,Gannouji:2018ncm,Basilakos:2019hlb}). In particular, recent observations indicate that $\gamma >0.55$ (weaker growth rate) in contrast to the usual theoretical prediction of $\gamma <0.55$  that is supported by many modified gravity models such as $f(R)$ theories \cite{Yin:2018mvu} and indicates stronger gravity at low $z$. In  what follows we focus on the parametrization \eqref{PKeq:geffansatz}.

The $\mu$ parametrization \eqref{PKeq:geffansatz} has been extensively studied in Refs. \cite{Nesseris:2017vor,Kazantzidis:2018rnb, Perivolaropoulos:2019vkb}, where it was shown that in the context of a wide range of different RSD datasets, a negative value of the parameter $g_a$ is favored in the context of {\plcdm background expansion rate $H(z)$ ($g_a=-0.68\pm 0.18$) indicating weaker gravity than the GR prediction at low $z$. 

This trend for weaker gravity at low $z$ is also supported by WL data \cite{Hildebrandt:2016iqg,Kohlinger:2017sxk,Joudaki:2017zdt,Abbott:2017wau,DiValentino:2018gcu} even though in these references this trend was expressed as a trend for lower values of $\sigma_8$ and \omom (or equvalently $S_8\equiv  \sigma_8 \sqrt{\Omega_{0m}/3}$) compared to the \plcdm best fit since $\mu$ was fixed to unity. This tension level which can not be released even by the inclusion of massive neutrinos, is demonstrated in Fig.  \ref{fig:S8const}. In Fig. \ref{fig:S8const} $A_{lens}$ is an effective parameter that rescales the lensing amplitude in the CMB spectra. The extension of \lcdm involving the parameter $A_{lens}$ is degenerate with a modified gravity extension and can clearly decrease the $\sigma_8$ tension implied by the WL data as shown in Fig. \ref{fig:S8const} \cite{DiValentino:2018gcu}. In contrast, the introduction of massive sterile neutrinos appears to have small effect on the tension level.

As discussed above, an evolving $\mu$ can also affect the $H_0$ tension problem. Indeed, local measurements of $H_0$ are heavily based on SnIa as distance indicators and on the assumption that after proper calibration the SnIa absolute magnitude $M$ may be assumed to be constant.  The peak luminosity of SnIa, is related to the gravitational constant as $L \propto G^{-3/2}$ \cite{Gaztanaga:2001fh}, which leads to an absolute magnitude $M$ that is associated with $\mu$ through \cite{Amendola:1999vu,Gaztanaga:2001fh}
\be 
M-M_0=\frac{15}{4} \, log_{10} \left(\mu \right)
\label{PKeq:mgeffconn}
\ee
where $M_0$ is a reference asymptotic value of the absolute magnitude.  A more detailed and accurate approach for determing the dependence of $M$ of SnIa on the Newton's constant has been implemented in Ref. \cite{Wright:2017rsu} through a semi-analytical method of light curve fitting which uses the standardised intrinsic luminosity L instead of  the peak luminosity of individual events to find the dependence of $M$ on the value of G. In this model the sign of the power index $-3/2$ appearing above, is indicated to be positive instead. This possibility will be discussed in a following publication. Usually, $M$ is considered to be a constant nuisance parameter and is marginalized. However, since dynamical probes favor a $\mu$ smaller than the GR value, similar trends (perhaps not of the same magnitude due to the strong gravitational fields involved) are expected for the absolute magnitude $M$. In what follows we present a short preliminary analysis attempting to address this issue and identify possible trends and constraints in the absolute $M$ of the SnIa.

In the context of the above discussion, the following questions arise:
\begin{itemize}
\item What is the current level of the $\fs$ tension and what is the implied evolution of $\mu$ in the context of \lcdmnospace?
\item Are there hints of a similar evolution of $\mu$ in the Pantheon SnIa dataset?
\item What is the allowed evolution of $\mu$ from the low $l$ CMB data?
\end{itemize}
These questions will be addressed in what follows.

The structure of this brief review is the following: In Sec. \ref{sec:fstensionmodgrav} we review the $\fs$ tension and the implications of a dynamical $\mu(z)$ for modified gravity theories. In Sec. \ref{sec:Pantheon} a tomographic analysis of the SnIa absolute magnitude of the Pantheon dataset is performed and the constraints on possible evolution at low $z$ are specified. Finally, in Sec. \ref{sec:ISW} the constraints on an evolving $\mu$ from the low $l$ angular CMB spectrum and the ISW effect are presented in the context of a \lcdm background, while in Sec. \ref{sec:Conclusions} we outline and discuss our results.

\section{The $\fs$ Tension and Modified Gravity.}
\label{sec:fstensionmodgrav}
\subsection{Observational Evidence}
The solution of eq. \eqref{PKeq:oddD} with initial conditions deep in the matter era, a $wCDM$ background \eqref{PKeq:wcdm} and an evolving parameter $\mu(z)$ of the form \eqref{PKeq:geffansatz} with $n=2$ respects both nucleosynthesis constraints and Solar System constraints. The theoretical prediction for $\fsz$ obtained from such a solution using also eq. \eqref{PK:eqfs8} depends on the parameters $\Omega_{0m}$, $w$ and $g_a$ and is shown in Fig. \ref{fig:fs8z}, along with a large compilation of corresponding datapoints \cite{Kazantzidis:2018rnb} (the different colors correspond to early or more recent time of publication). Clearly, the parameter values ($\Omega_{0m}, w,\sigma_8, g_a)=(0.31,-1,0.83,0)$ corresponding to \plcdm  lead to larger growth ($\fsz$) than most data would imply, especially at redshifts $z<1$ (red line). The fit to the data may be improved either by modifying the background expansion rate $H(z)$ (\eg lowering $\Omega_{0m}$) and/or by lowering the strength of gravity at low $z$. Fixing the background $H(z)$ to \plcdm and allowing $g_a$ in eq. \eqref{PKeq:geffansatz} to vary we obtain \cite{Nesseris:2017vor,Kazantzidis:2018rnb} a best fit value of $g_a=-0.68\pm 0.18$ for $n=2$ which is approximately $3.7\sigma$ away from the GR value $g_a=0$.

\begin{figure}[!h]
\centering
\includegraphics[width = 0.5\textwidth]{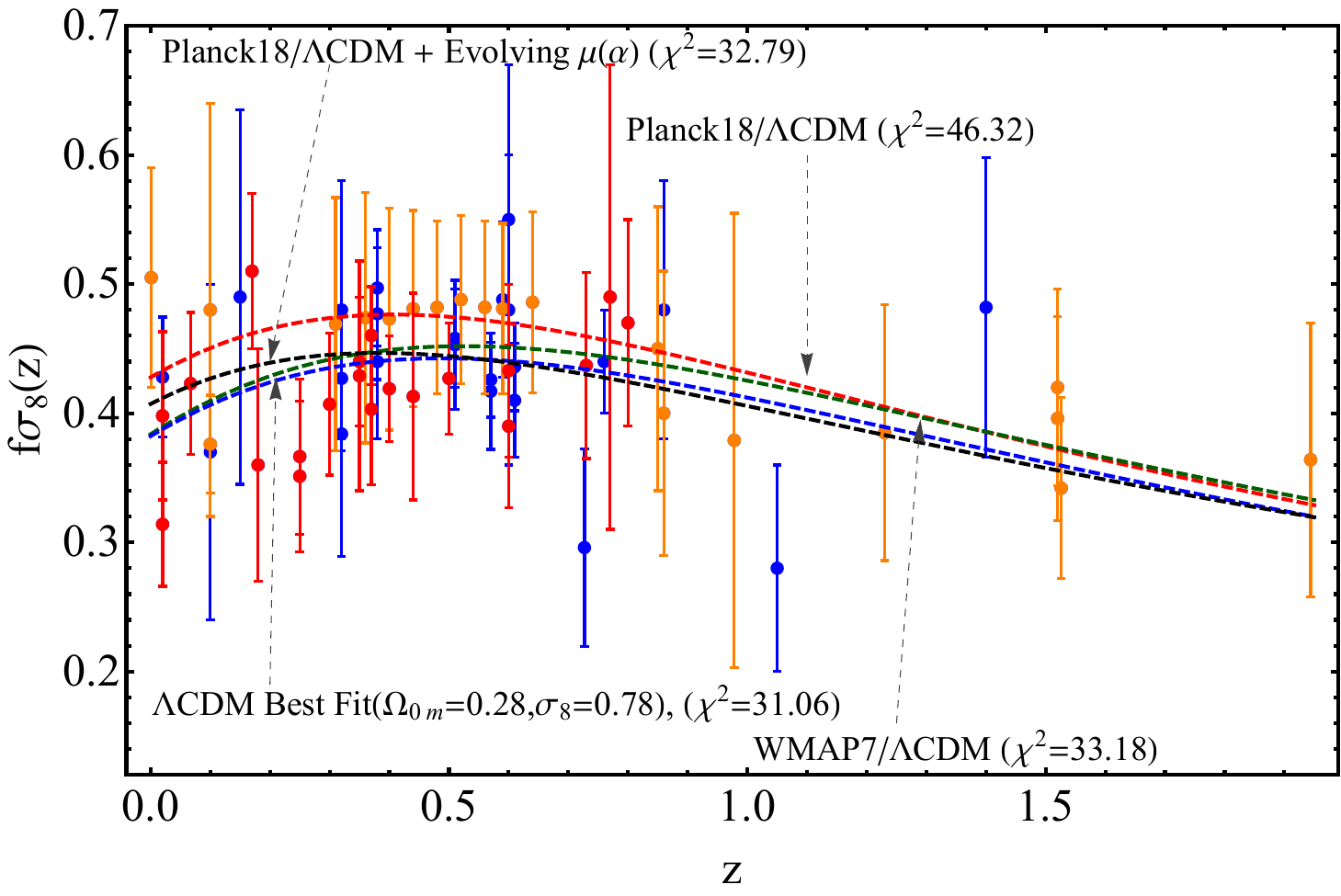}
\caption{Evolution of $\fs$ as a function of redshift. The red dashed line corresponds to the \plcdm model ($\Omega_{0m}=0.315 \pm 0.007, \sigma_8= 0.811 \pm 0.006$), the green one to the \wlcdm ($\Omega_{0m}=0.266 \pm 0.025, \sigma_8= 0.801 \pm 0.030$), the black one to an evolving $\mu$ with a \plcdm background ($\Omega_{0m}=0.315 \pm 0.007, \sigma_8= 0.811 \pm 0.006$, $g_a =-0.681 \pm 0.177$), while the blue one describes the best fit \lcdm coming from the 63 compilation of Ref. \cite{Kazantzidis:2018rnb} ($\Omega_{0m}=0.279 \pm 0.028, \sigma_8= 0.775 \pm 0.018$). The orange points correspond to the 20 latest datapoints, while the red ones to the 20 earliest of this compilation. The blue points account for the rest of the growth data.}
\label{fig:fs8z}
\end{figure}

The trend for weaker gravity at low redshifts is also evident in Fig. \ref{fig:Geffaplot} which shows the best fit form of $\mu(a)$ as a function of the scale factor $a$ for the best fit values of $g_a$ coming from the robust RSD data compilation of Ref. \cite{Nesseris:2017vor} for different values of $n$. The required drop of $\mu(a)$ becomes stronger and localized to low $z$ as $n$ increases. As discussed in Sec. \ref{sec:ISW}, however, such a large drop is not consistent with the low $l$ CMB angular power spectrum and the ISW effect. 

\begin{figure}[!h]
\centering
\includegraphics[width = 0.5\textwidth]{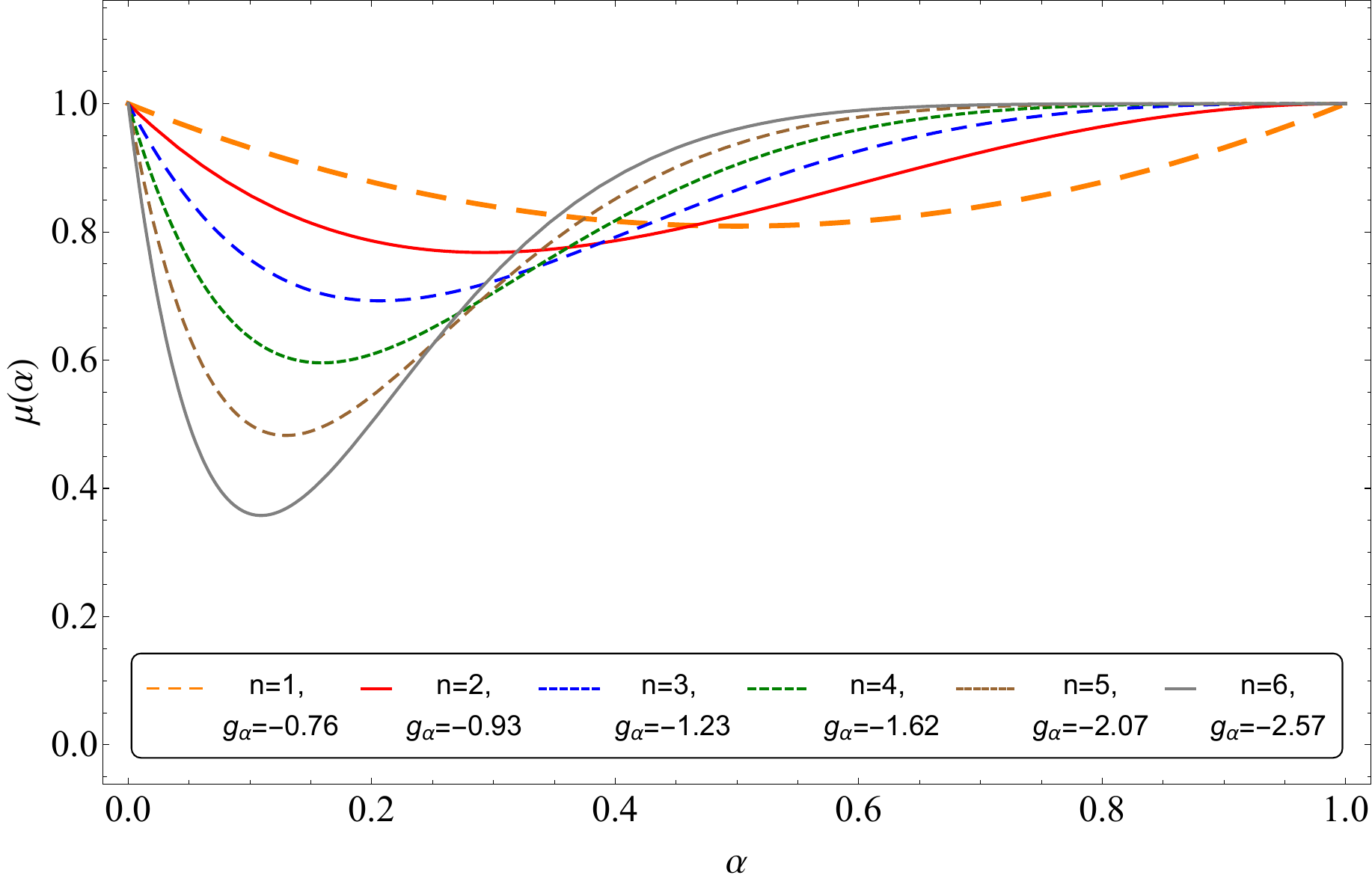}
\caption{Evolution of $\mu$ as a function of the scale factor $a$ considering the best fit values for $g_a$ and various values of $n$ using the robust collection of Ref. \cite{Nesseris:2017vor}}.
\label{fig:Geffaplot}
\end{figure}

An interesting feature of the theoretical model predictions for $\fsz$ shown in Fig. \ref{fig:fs8z} is the degeneracy among these predictions for $z>1$. This degeneracy has been investigated in some detail in \cite{Kazantzidis:2018jtb} for $\fsz$ and for other cosmological observables. It was found that there are blind redshift spots where observables are degenerate with respect to specific cosmological parameters. For $\fsz$ with respect to the parameter $g_a$ there is a blind spot at $z\simeq 2.5$ and its constraining power is significantly reduced for $z>1$. Thus $\fsz$ datapoints with $z<1$ can constrain $\mu(z)$ (or equivalently $g_a$) much more efficiently than points at higher redshifts. This is demonstrated in Fig. \ref{fig:deltafs8}  which shows the difference between the growth rate in the context of an evolving $\fsz$ from the \plcdm $\fsz$ \cite{Kazantzidis:2018jtb}  for various values of $g_a$. This difference is defined as
 
\begin{figure}[!h]
\centering
\includegraphics[width = 1.\textwidth]{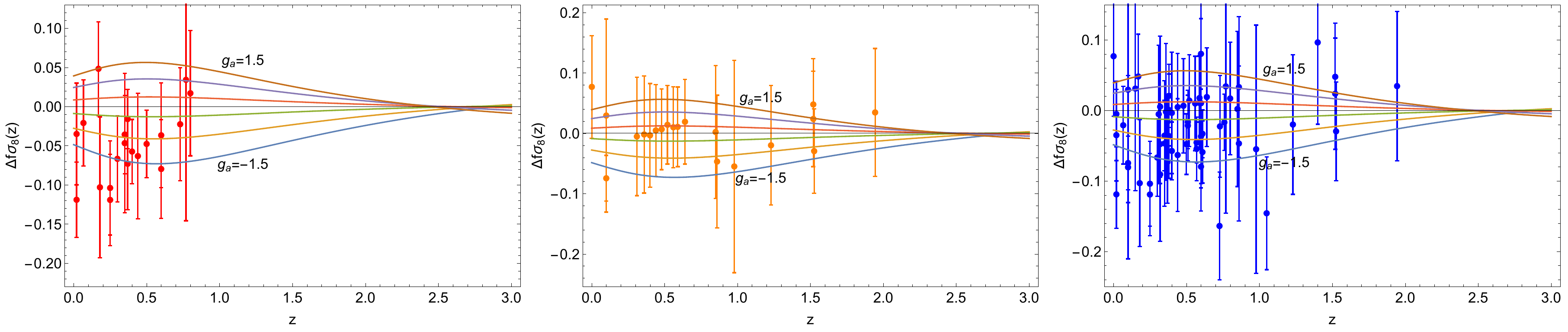}
\caption{Evolution of $\Delta \fs$ as a function of the redshift $z$ for different values of $g_a$. These curves are superimposed with the 20 earliest datapoints (left panel), the 20 latest (middle panel) and the full compilation (right panel) of Ref. \cite{Kazantzidis:2018rnb}. Notice that early lower $z$ datapoints are much more efficient in detecting hints of modified gravity (a non-zero value of $g_a$).}
\label{fig:deltafs8}
\end{figure}

\be 
\Delta \fs= \fs (z,\Omega_{0m}^{Planck18},-1,g_a) -\fs (z,\Omega_{0m}^{Planck18},-1,0)
\ee
Clearly, early published datapoints which tend to have lower redshifts (right panel) have more constraining power than more recently published datapoints (middle panel) which have higher $z$ and larger errorbars. The tension level comes mainly from early datapoints which appear to favor $\Delta \fs<0$ \ie weaker growth. \footnote{The RSD datapoints  of Fig. \ref{fig:deltafs8} include a $1-3\%$ ``fiducial cosmology" Alcock-Paczynski correction \cite{Alcock:1979mp,Macaulay:2013swa,Kazantzidis:2018rnb,Nesseris:2017vor} \ie they have been multiplied by a factor $\frac{H(z) d_A(z)}{H_{fid}(z)d_{A_{fid}}(z)}$ where the subscript $_{fid}$ indicates the  fiducial cosmology used in each survey to convert angles and redshift to distances for evaluating the correlation function and $H(z), d_A(z)$ correspond to the Hubble parameter and the angular diameter distance of the true cosmology.}

The trend for weaker growth of matter perturbation than the growth favored by \plcdm has been pointed out in a wide range of studies in the context of different dynamical probe data. One of the first analyses that pointed out the weak growth tension was that of Ref. \cite{Macaulay:2013swa} where it was pointed out that RSD measurements are consistently lower than the values expected from Planck in the context of \lcdmnospace. It was also pointed out that other dynamical probes like the Sunyaev-Zeldovich (SZ) cluster counts \cite{Ade:2013lmv} also indicate weaker growth ($\sigma_8=0.77\pm 0.02$, $\Omega_{0m}=0.29\pm 0.02$). Similar trends were found earlier, using the measurement of the galaxy cluster cluster mass function in the redshift range $z\in [0,0.9]$ \cite{Vikhlinin:2008ym}, where lower values of $\Omega_{0m}$ and $\sigma_8$ were favored. Later studies confirmed this trend by pointing out that best fit cosmological parameters like the matter density $\Omega_{0m}$ and the dark energy equation of state $w$ differ at a level of $2-3\sigma$ between geometric probes (SnIa, BAO and CMB peak locations) and dynamical probes (RSD data, CC and WL) \cite{Ruiz:2014hma,Bernal:2015zom}. The dynamical probes of growth pointed consistently towards lower values of $\Omega_{0m}$ and thus weaker growth. It was also realized that in particular, WL data indicated consistently a $2-3\sigma$ tension with the Planck parameter values of $\Omega_{0m}-\sigma_8$ \cite{Ade:2015xua,Joudaki:2017zdt,DiValentino:2018gcu} (for updated constraints see also Fig. \ref{fig:S8const} adopted from Ref. \cite{DiValentino:2018gcu}). For example, the Kilo-Degree Survey (KiDs-450) \cite{Joudaki:2017zdt,Joudaki:2016kym} finds $S_8\equiv \sigma_8 \sqrt{\Omega_{0m}/0.3}=0.74\pm 0.035$ which is smaller at a $2.6\sigma$ tension compared to the corresponding Planck best fit value  $S_8=0.832\pm 0.013$ \cite{Aghanim:2018eyx}. More recent WL cosmic shear data from the Dark Energy Survey \cite{Troxel:2017xyo,Abbott:2017wau} indicate $S_8=0.792\pm 0.024$, \ie a weaker tension with geometric probes and Planck (about $1-2\sigma$), albeit in the same direction of weaker growth and lower  $\Omega_{0m}-\sigma_8$ (DES indicates that $\Omega_{0m}=0.264^{+0.032}_{-0.019}$ \cite{Abbott:2017wau} to be compared with Planck best fit $\Omega_{0m}=0.315\pm0.007$  \cite{Aghanim:2018eyx}). Reduced value of $\sigma_8$ ($\sigma_8=0.77\pm 0.02$) is also indicated by high $l$ measurements ($l>2000$) of the E-mode angular auto-power spectrum (EE) and the temperature-E-mode cross-power spectrum (TE)  taken with the SPTpol instrument \cite{Henning:2017nuy}.

The tension level between geometric and dynamical probes has recently been quantified by using specific statistics designed to probe the tension in a more efficient and quantitative manner \cite{Raveri:2015maa,Lin:2017ikq,Sagredo:2018rvc}.  These studies have verified the statistical significance of the tension between geometric and dynamical probes and demonstrated that even though the dynamical probes (RSD, WL and CC) are consistent with each other pointing towards weaker growth than GR, they are in discordance with the geometric probes in the context of GR. 

\subsection{Theoretical Implications}
The most generic approach to the ``weak growth" tension is the modified gravity approach. If this tension is in fact due to a modification of GR on cosmological scales the following question arises: ``What observationally viable modified gravity models can reproduce a weaker gravity than that predicted by GR at low redshifts?" A naive response to this question would indicate that any viable modified gravity model can lead to weaker gravity than GR at late times with proper choice of its parameters. However, it may be shown that this is not the case. Recent studies have addressed this question for $f(R)$ theories, for minimal scalar tensor theories  \cite{Gannouji:2018ncm,Perivolaropoulos:2019vkb} for Horndeski theories  \cite{Arjona:2019rfn,Linder:2018jil} and beyond Horndeski  Gleyzes-Langlois-Piazza-Vernizzi (GLPV) theories \cite{Tsujikawa:2015mga}.

We will consider $f(R)$ models with an action of the form
\be
S=\int d^4x \sqrt{-g} \, \frac{f(R)}{2}+ S_m,\label{PKeq:mogaction}
\ee
where from now on we set $8\pi G_{\textrm{N}}=1$. The predicted $\mu(z,k)$ is given as \cite{Tsujikawa:2007gd}
\be 
\mu(z,k)=\left(\frac{d f}{dR} \right)^{-1} 
\left[ \frac{1+4 \left(\frac{ d^2 f}{d R^2}/ \frac{ d f}{d R}\right) \cdot k^2 \, 
(1+z)^2}{1+3 \left(\frac{ d^2 f}{d R^2}/ \frac{ d f}{d R}\right) \cdot k^2 \, (1+z)^2} \right] \label{PKeq:gefffr}
\ee
where in this case $\mu$ depends on both the redshift $z$ and the scale $k$. In addition, the stability conditions
\ba 
\frac{ d^2 f}{d R^2}&>&0 \nn \\
\frac{ d f}{d R}&>&0
\ea
should be satisfied \cite{Starobinsky:2007hu}. Also in viable $f(R)$ models $\frac{d f}{d R} \simeq 1$ at early times deep in the matter era (high $R$) \cite{Amendola:2006we}. Thus, since $\frac{ d^2 f}{d R^2}>0$ we must have $\frac{ d f}{d R}<1$ at late times (low $R$). It follows that both factors of eq. \eqref{PKeq:gefffr} are larger than unity and we have generically in $f(R)$ theories that $\mu(z)\geq 1$. This is a generic result independent of the background $H(z)$, indicating that $f(R)$ theories are unable to resolve the weak growth tension because they predict stronger gravity than GR.

A similar result is true for minimal scalar-tensor theories, provided that the expansion background is close to \lcdmnospace. The minimal scalar-tensor action has the form \cite{Boisseau:2000pr}
\be
S= \int d^4 x \sqrt{-g} \left[ \frac{1}{2}F(\phi)R - \frac{1}{2} g^{\mu\nu} \partial_\mu \phi \partial_\nu \phi - U(\phi) \right] + S_m,
\label{PKeq:actionscalten}
\ee
The dynamical equations obtained by variation of this action in the context of a flat FRW metric are of the form 
\cite{EspositoFarese:2000ij,Boisseau:2000pr}
\begin{eqnarray}
3F H^2 &=&  \rho +{\frac{1}{2}} \dot\phi^2 - 3 H \dot F + U \label{PKeq:fe1}\\
-2F \dot H  &=& (\rho+p) + \dot \phi^2 +\ddot F - H \dot F \label{PKeq:fe2} 
\end{eqnarray}
where the dot represents differentiation with respect to cosmic time $t$. After rewriting the equations of motion in terms of the redshift, defining the rescaled square Hubble parameter as $q(z)=\frac{H^2(z)}{H_0^2}$  and eliminating the scalar field potential $U(\phi)$, we obtain a differential equation that associates the coupling function $F(\phi)$ and the scalar field $\phi$ as
\be F^{\prime\prime}(z)+ \left[\frac{q^\prime(z)}{2q(z)}-\frac{2}{1+z}\right] F^{\prime}(z)  - \frac{1}{(1+z)}\frac{q^\prime(z)}{q(z)} F(z)+3 \frac{1+z}{q(z)} \Omega_{0m}= -\phi^{\prime}(z)^2 \label{PKeq:fe1a} 
\ee
where the prime stands for differentiation with respect to redshift $z$.

In scalar tensor theories $\mu$ is expressed as \cite{EspositoFarese:2000ij,Nesseris:2006jc}
\be
\mu(z)= \frac{1}{F(z)}\frac{F(z)+2F_{,\phi}^2}
{F(z)+\frac{3}{2}F_{,\phi}^2}
\label{PK:eqgeffsclatens}
\ee
Using eq. \eqref{PK:eqgeffsclatens} in the differential equation \eqref{PKeq:fe1a} and expanding around $z=0$, while using the Solar System constraint $\mu'(z=0)=0$ \cite{Gannouji:2006jm,Nesseris:2006hp} we find in the context of a $wCDM$ background  \cite{Gannouji:2018ncm}

\be 
\mu''(0)=9(1+w)(-1+\Omega_{0m})+\frac{9(1+w)^2 (-1+\Omega_{0m})^2}{\phi'(0)^2} +2\phi'(0)^2  \label{PKeq:geffdp0sct}
\ee
For a \lcdm background, eq. \eqref{PKeq:geffdp0sct} leads to the low $z$ expansion
\be 
\mu(z) \approx  \mu(0)+ \frac{1}{2} \mu''(0)z^2=1+\phi^{\prime}(0)^2 \, z^2+\ldots \label{eq:geffform}
\ee
which implies that $\mu(z)$ can only increase with redshift around $z=0$ in the context of a \lcdm backround. In fact this result ($\mu''(0)>0$) is also applicable for $w<-1$ as it can be seen from eq. \eqref{PKeq:geffdp0sct}, while for $w>-1$ it is possible to have $\mu''(0)<0$ as shown in Fig. \ref{fig:Geffdoubprime}. 

\begin{figure}[h]
\centering
\includegraphics[width = 0.5\textwidth]{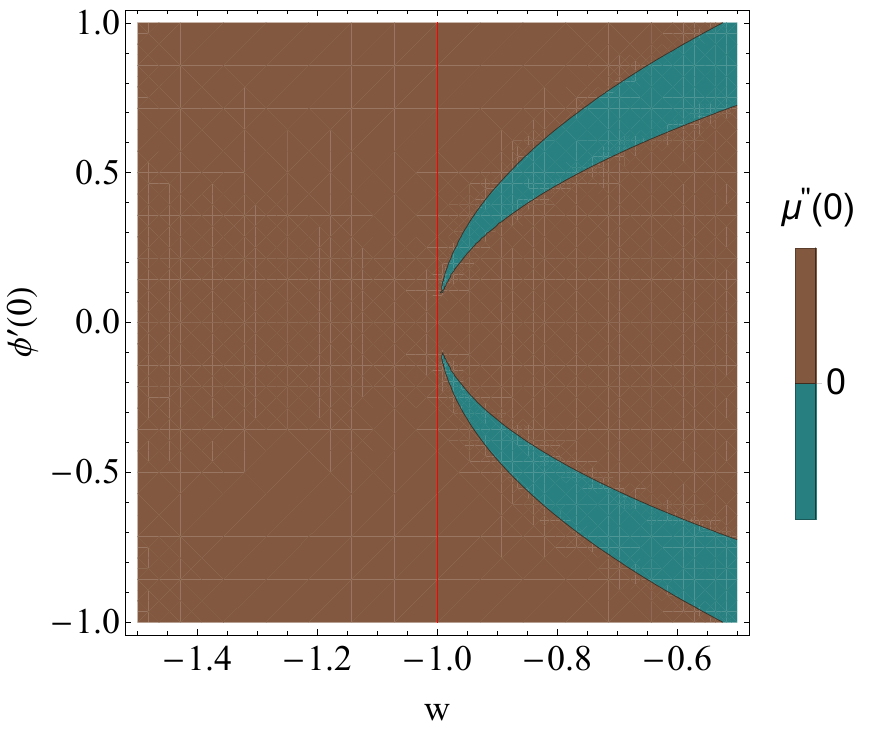}
\caption{The second derivative of $\mu(0)$ in the parametric space of $\phi^{\prime}(0)$ and $w$. The brown region describes the parameter values for $\mu''(0)>0$, while the blue region describes the parameter values for $\mu''(0)<0$ (from Ref. \cite{Gannouji:2018ncm}).}
\label{fig:Geffdoubprime}
\end{figure}

\noindent Thus, the increasing nature of $\mu(z)$ in scalar tensor theories that respect the solar system constraints has been demonstrated analytically in the context of a \lcdm background around $z=0$. 
\begin{figure}[!h]
\centering
\includegraphics[width = 0.5\textwidth]{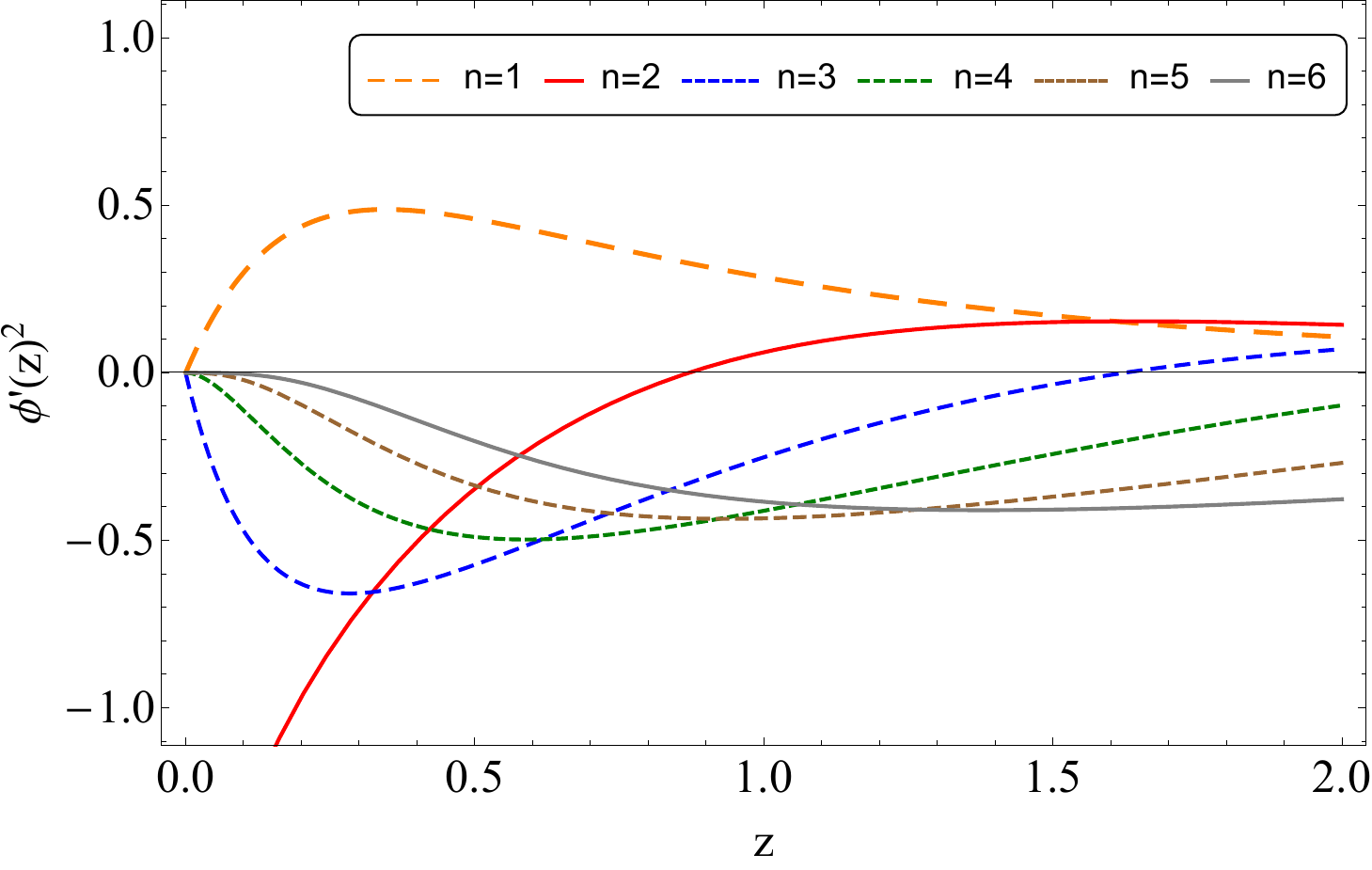}
\caption{Evolution of the scalar field $\phi$ as a function of redshift $z$ corresponding to the best fit values of $g_a$ and various values $n$ using the robust compilation of Ref. \cite{Nesseris:2017vor}. Notice that the $n=1$ case shows no ghost instabilities ($\phi'(z\simeq 0)>0$) but it does not satisfy the solar system constraint $\mu'(z=0)=0$) and thus
 eq. (\ref{eq:geffform}) is not applicable for $n=1$.}
\label{fig:phizplot}
\end{figure}

This result is also demonstrated numerically by using Eqs. \eqref{PKeq:fe1a}, \eqref{PK:eqgeffsclatens} in the context of the best fit parametrization \eqref{PKeq:geffansatz} obtained from the RSD growth data \cite{Nesseris:2017vor} ($g_a<0$). Fig. \ref{fig:phizplot} shows the corresponding evolution of $\phi'(z)^2$, demonstrating that, as expected, for a decreasing $\mu(z)<1$ we obtain $\phi'(z)^2<0$ (ghost instabilities) at least close to $z=0$ when the Solar System constrants are respected (this does not include the $n=1$ case).  In the case of more general scalar-tensor theories (Horndeski and beyond Horndeski) it has been shown that weaker gravity may be possible provided specific constraints among the terms of the Lagrangian are applicable \cite{Tsujikawa:2015mga,Linder:2018jil}. 
  
We therefore conclude that from the theoretical point of view it is highly challenging to construct a viable theoretical model that allows for weaker gravity than GR at low redshifts while at the same time it respects solar system and other observational constraints with an $H(z)$ background close to \lcdmnospace. This challenge, however, may prove a useful discriminating tool among modified gravity models if the weak growth tension persists and gets verified by future cosmological data.  

The issue of weak growth tension is expected to be clarified within the next decade due to a wide range of  upcoming surveys. The surveys include Euclid \cite{Laureijs:2011gra,Amendola:2016saw} (aiming at mapping the geometry of the Universe when dark energy lead to its accelerated expansion), Square Kilometer Array (SKA) \cite{Jarvis:2015tqa,Bacon:2015dqe}  (aiming at analyzing radiosignals from various galactic sources), Large Synoptic Survey Telescope (LSST) \cite{Marshall:2017wph} (aiming at mapping and cataloging galaxies, in order to study their impact on the distortion of spacetime), Cosmic Origins Explorer (COrE) (aiming at mapping the polarization of the CMB) \cite{Bouchet:2015arn},  Dark Energy Spectroscopic Instrument (DESI) \cite{Aghamousa:2016zmz,Aghamousa:2016sne} (aiming at studying the effects of dark energy and obtaining the optical spectra of galaxies and quasars) and Wide Field Infrared Survey Telescope  (WFIRST) \cite{Spergel:2015sza,Hounsell:2017ejq} (aiming at answering key questions in cosmology, probing BAO, WL and Supernovae data simultaneously). These surveys are expected to provide new more detailed measurements of the dark energy probes BAO, SnIa, RSD, WL and CC extending to both dynamical and geometrical probes. They are expected to either confirm or eliminate the weak growth tension. In the first case, they will also provide a concrete discriminator among the modified gravity models and non-gravitational models that constitute candidate extensions of the standard \lcdm model and are motivated by the weak growth tension.

\section{Evolving $\Geff$ and the Pantheon SnIa dataset.}
\label{sec:Pantheon}

If the effective Newton's constant $\mu=\Geff/G_{\textrm{N}}$ is indeed evolving with redshift on cosmological timescales it is expected to lead to an evolution of the absolute luminosity and absolute magnitude of SnIa. In this section we present preliminary work searching for such evolution of the  SnIa absolute magnitude with redshift. We use the Pantheon SnIa dataset \cite{Scolnic:2017caz}, which is the latest compilation of SnIa. It consists of 1048 data with redshifts spanning the region $z \in \left[0.01,2.3\right]$.  This dataset is a combined set of the PS1 SnIa dataset \cite{Scolnic:2013efb}, which consists of 279 SnIa with redshifts spanned in the region $z \in \left[0.03,0.68\right]$ along with 
probes of low redshifts ($z \in \left[0.01,0.1\right]$), including the  CfA1-CfA4 \cite{Riess:1998dv,Jha:2005jg,Hicken:2009df,Hicken:2012zr} and CSP surveys \cite{Contreras:2009nt,Stritzinger:2011qd}, as well as high redshifts ($z >0.1$), probed by SDSS \cite{Sako:2014qmj,Kessler:2009ys}, SNLS \cite{Sullivan:2011kv,Conley:2011ku} and HST surveys \cite{Riess:2006fw,Suzuki:2011hu}. 

The measured apparent magnitude $m$ for SnIa data is connected to cosmological parameters through the relation 
\be 
m_{th}(z)=M+5 \, log_{10} \left[\frac{d_L(z)}{Mpc} \right]+25 \label{PKeq:appmagn}
\ee 
where $d_L(z)$ is the  luminosity distances and $M$ is the absolute magnitude. The luminosity distance for a flat FLRW metric, is given by 
\be 
d_L=c(1+z) \int_0^{z} \frac{dz'}{H(z')}
\ee

The Pantheon dataset provides the apparent magnitude $m_{obs}(z_i)$ after corrections over the stretch, color and possible biases from simulations \cite{Scolnic:2017caz}. Following the usual method of maximum likelihood  \cite{Arjona:2018jhh}  we can obtain the best fit parameters, minimizing the quantity
\be 
\chi^2 (M,\Omega_{0m},w,h)=V^i_{Panth.} C_{ij}^{-1} V^j_{Panth.}
\ee
where $V^i_{Panth.}\equiv m_{obs}(z_i)-m_{th}(z)$, $C_{ij}$ provided in \cite{Scolnic:2017caz}, is the covariance matrix and $h$ is the dimensionless parameter of the Hubble constant, which is defined as $h \equiv H_0/100 \, (km/s)/Mpc$. 

Usually, the absolute magnitude $M$ is considered a nuisance parameter and is marginalised along with $h$, due to a clear degeneracy between the two parameters. However, in the context of modified gravity with an evolving Newton's constant the absolute magnitude is expected to evolve with redshift in accordance with eq. \eqref{PKeq:mgeffconn} and may contain useful information on fundamental physics. In an effort to identify such evolution we minimize $\chi^2$ with respect to the parameter $M$ with fixed background corresponding to the best fit \lcdm $H(z)$ as obtained from the full Pantheon dataset with $M$ marginalization (we fix $w=-1$ and $\Omega_{0m}=0.28$ \cite{Scolnic:2013efb} and set $h=1$ for simplicity). In this context, we identify the best fit value and $1\sigma$ error of $M$ for various subsets of the full Pantheon dataset.
In Fig. \ref{fig:erroplotMcut} we show the best fit absolute magnitude $M$  using subsamples of the Pantheon dataset in the redshift range $z \in \left[0.01, z_{max}\right]$. The $1 \sigma$ range for the $M$ parameter for various cutoffs $z_{max}$ is also shown.

\begin{figure}[!h]
\centering
\includegraphics[width = 0.48\textwidth]{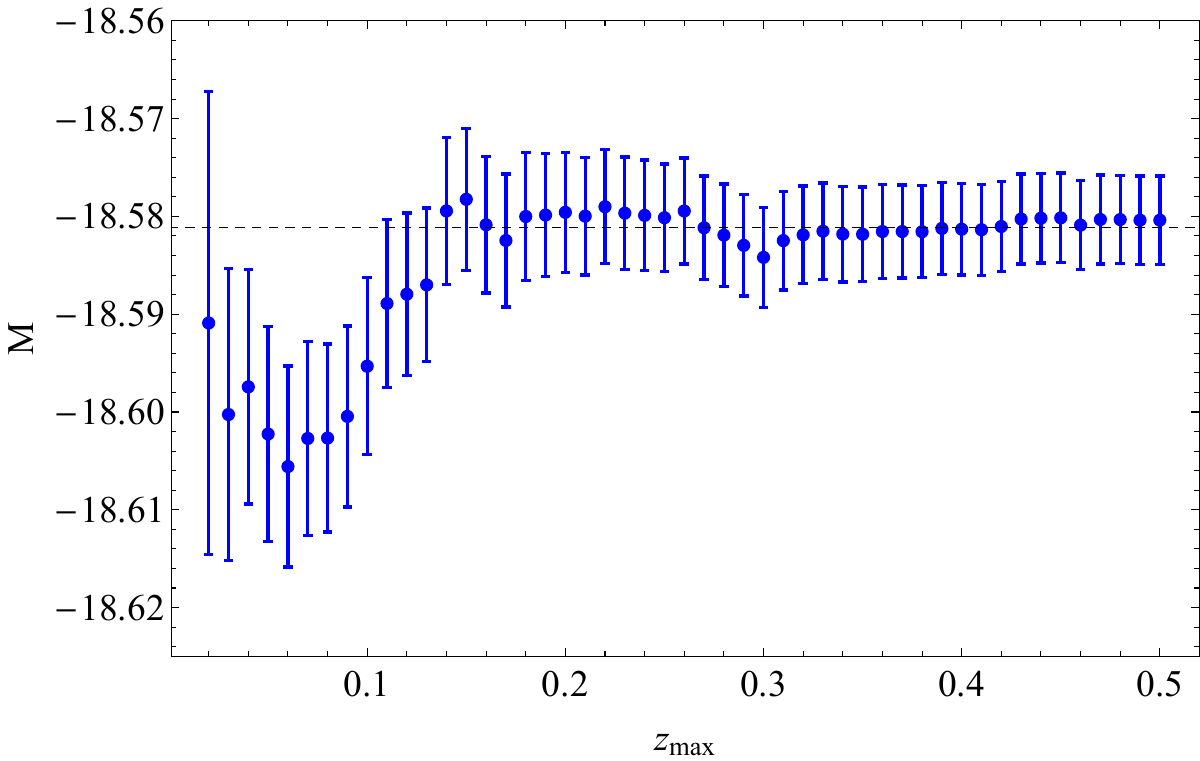}
\caption{Evolution of the absolute magnitude $M$ as a function of the cutoff $z_{max.}$. We have set $h=1$ and thus the value of $M$ is shifted compared to its usual value of $M=-19.3$.}
\label{fig:erroplotMcut}
\end{figure}

It is clear from Fig. \ref{fig:erroplotMcut} that low resdhift data in the redshift range $z \in \left[0.01, 0.1\right]$ seem to favour a value $M$ smaller than its best fit asymptotic value based on the full dataset ($z_{max}=2.3$) at a level of about $2\sigma$. At redshifts $z>0.2$, $M$ approaches its asymptotic value (dashed line). Our results are consistent with the analysis of Ref. \cite{Colgain:2019pck} where the best fit parameters of $\Omega_{0m}$ and $H_0$ were investigated as a function of the redshift cutoff $z_{max}$. In agreement with our results it was found that the low $z$ Pantheon data appear to have interesting features which may indicate the presence of either systematics or new physics.

In Fig. \ref{fig:mu} (left panel) we show the 100 point moving best fit value of $M$ along with its $1\sigma$ errors. To construct this plot we rank the Pantheon datapoints from lowest to highest redshift. We start  with the first 100 datapoints (lowest redshift points 1 to 100) and use them to obtain the best fit value of $M$ (assuming the fixed \lcdm background) with its $1\sigma$ error. The corresponding $z$ coordinate of this point is the mean redshift of the first 100 points. The $i^{th}$ point is obtained by repeating the above procedure for the datapoints from $i$ to $i+100$.

Using the best fit value of $M$ obtained from each 100 point subsample we can calculate $\mu$ using eq. \eqref{PKeq:mgeffconn} (right panel of Fig. \ref{fig:mu}) setting $M_0$ equal to the best fit value of $M$ obtained from the full Pantheon dataset. Clearly, an oscillating effect is evident for the absolute magnitude $M$ at low $z$, that is eased at high redshifts. The same oscillating effect was also observed in Refs. \cite{Kazantzidis:2020tko,Sapone:2020wwz,Kazantzidis:2020xta}, where a binning method was used instead.

\begin{figure}[h]
\centering
\includegraphics[width = 0.8\textwidth]{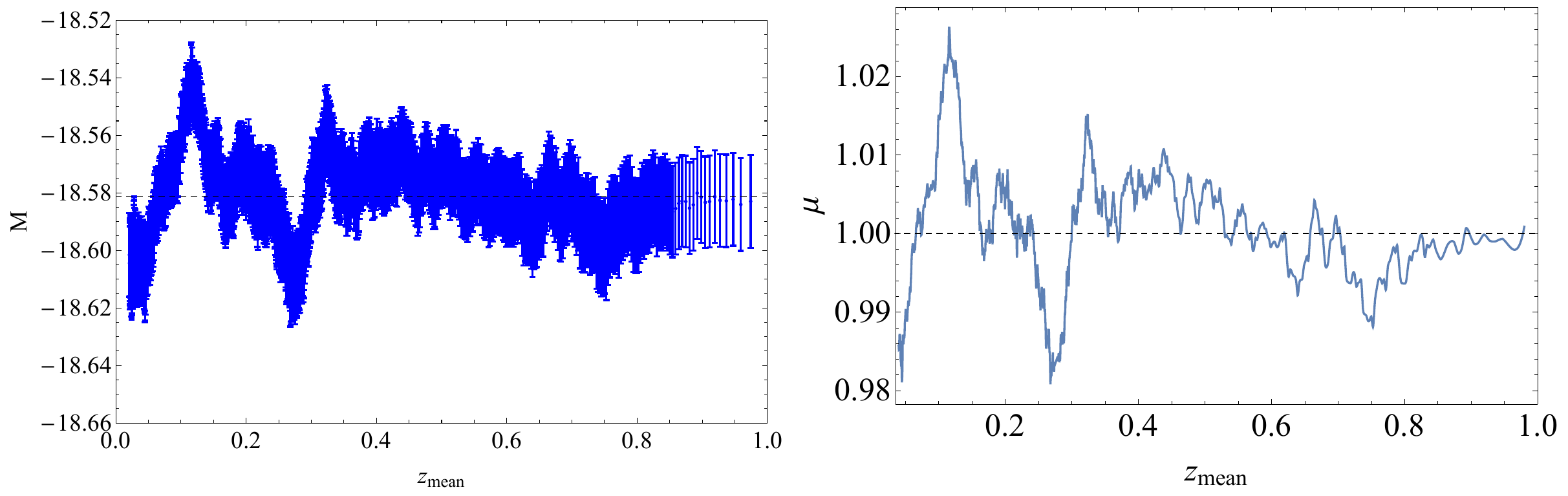}
\caption{Left Panel: Variation of the absolute magnitude $M$ as a function of $z_{mean}$ for 100 point Pantheon subsamples. Right Panel: The corresponding variation of $\mu$ as a function of $z_{mean}$ for 100 datapoint Pantheon subsamples.}
\label{fig:mu}
\end{figure}

The absolute magnitude $M$ is degenerate with $h$. Therefore, the value of $h$ obtained under the assumption of a constant $M$ would not be the same as the value of $h$ that would be obtained if $M$ was allowed to evolve. In particular, using the ``Hubble constant free" luminosity distance, that is defined as \cite{Nesseris:2005ur} 
\be 
D_L(z)=\frac{H_0 \, d_L(z)}{c}
\ee
we can rewrite  eq. \eqref{PKeq:appmagn} as \cite{Nesseris:2005ur}
\be 
m_{th}(z)=M+5 \, log_{10} \left(D_L (z)\right)+5 \, log_{10} \left(\frac{c/H_0}{1 Mpc} \right)+25 \label{PKeq:appmagndL}
\ee
In terms of $h$ and taking into account the possible evolution of $\mu(z)$\footnote{The $\mu$ here is the evolving normalized Newton's constant and should not be confused with the distance modulus.}, eq. \eqref{PKeq:appmagndL} takes the form 
\be 
m_{th}(z)=M_0+\frac{15}{4} \, log_{10} \left(\mu (z) \right)+5 \, log_{10} \left(D_L (z)\right)-5 \, log_{10} \left(h\right)+42.38  \label{PKeq:appmagndLh}
\ee
Using eq. \eqref{PKeq:appmagndLh} it is easy to show that a change of $\mu$ by a small amount $\Delta \mu$ around $\mu=1$ is equivalent to a small change of $h$ by 
\be 
\Delta h= -\frac{3}{4}\, h\, \Delta\mu
\label{deltah}
\ee 
Thus, a decrease of $\mu$ at low $z$ ($\Delta\mu<0$) is equivalent to an increase of $h$ by $\Delta h$ compared to the true value of $h$. The value of $\Delta \mu \simeq -2\times 10^{-2}$ indicated in Fig. \ref{fig:mu} could be interpreted as to a shift of $h$ by about $1.5\%$ if $\mu$ was assumed fixed to 1. This artificial increase of $h$ is in the right direction but does not appear to be enough to explain a tension of about $8\%$ between the value indicated by the CMB and the value indicated by the SnIa sample. We stress, however, that the above analysis is heuristic and a more detailed analysis is required to include the possible effects of a varying $\mu$ in the derivation of $H_0$ from SnIa. In particular, the effects on Cepheid period-luminosity relation used in the determination of $H_0$ have not been taken into account, the effects of strong gravity in the interior of the progenitor stars have been ignored and the background cosmology has been assumed fixed to \lcdmnospace. These effects should be taken into account in a more complete and detailed analysis.

\section{Constraints on Evolving $\Geff$ from low $l$ CMB spectrum and the ISW effect.}
\label{sec:ISW}

As stated in the previous sections an evolving Newton's constant $\mu(z)$ would help resolve the weak growth and the $H_0$ tensions. However, such an evolution would also affect \cite{Giannantonio:2009gi} other dynamical probes and in particular the low $l$ (large scale) CMB angular power spectrum through the Integrated-Sachs Wolfe (ISW) effect created as the CMB photons travel through  time varying gravitational potential which would be modified by the evolving $\mu(z)$. Any such  modification is constrained by the Planck data. The questions that we address in this section is the following: 
\begin{itemize}
\item
What are the constraints imposed by the Planck CMB TT power spectrum data on the parameter $g_a$ of the parametrization \eqref{PKeq:geffansatz} assuming a fixed slip parameter to its GR value $\eta=1$?
\item
Are these constraints consistent with the value of $g_a$ required to resolve the weak growth tension?
\end{itemize}
In order to address these questions  we use the 2019 version \cite{Zucca:2019xhg} of MGCAMB \cite{Hojjati:2011ix,Zhao:2008bn}, which is a modified version of the CAMB code \cite{Lewis:1999bs}, that it is designed to produce the CMB spectrum in the context of modified gravity theories with a given background model $H(z)$ and a given scale dependent evolution of $\mu$ and $\eta$. We fix $H(z)$ to Planck15/$\Lambda$CDM, since the Planck18 likelihood chains which are implemented in COSMOMC and MGCOSMOMC are not yet publicly available, $\eta=1$ and for $\mu(z)$ we use the parametrization  \eqref{PKeq:geffansatz}. The values of the parameters for the Planck15/$\Lambda$CDM model are shown in Table  \ref{PKtab:plcdm15}. The predicted form of the CMB angular power spectrum for various values of $g_a$ is shown in Fig. \ref{fig:mgcamb} along with the corresponding Planck datapoints. 

\begin{table}[h!]
\caption{Planck15/$\Lambda$CDM  parameters values from Ref. \cite{Ade:2015xua} based on TT,TE,EE and lowP likelihoods. Notice that $\sigma_8$ is larger for the $2015$ data release which implies a stronger $\sigma_8$ tension that the Planck18/$\Lambda$CDM best fit model.}
\label{PKtab:plcdm15}
\begin{centering}
\begin{tabular}{cc}
 \hline 
 \rule{0pt}{3ex}  
  Parameter & Planck15/$\Lambda$CDM \cite{Ade:2015xua}\\
    \hline
    \rule{0pt}{3ex}  
$\Omega_b h^2$ & $0.02225\pm0.00016$ \\
$\Omega_c h^2$ & $0.1198\pm0.0015$  \\
$n_s$ & $0.9645\pm0.0049$\\
$H_0$ & $67.27\pm0.66$ \\
$\Omega_{0m}$ & $0.3156\pm0.0091$ \\
$w$ & $-1$  \\
$\sigma_8$ & $0.831\pm0.013$ \\
\hline
\end{tabular}
\end{centering}
\end{table}

\begin{figure}[h]
\centering
\includegraphics[width = 0.5\textwidth]{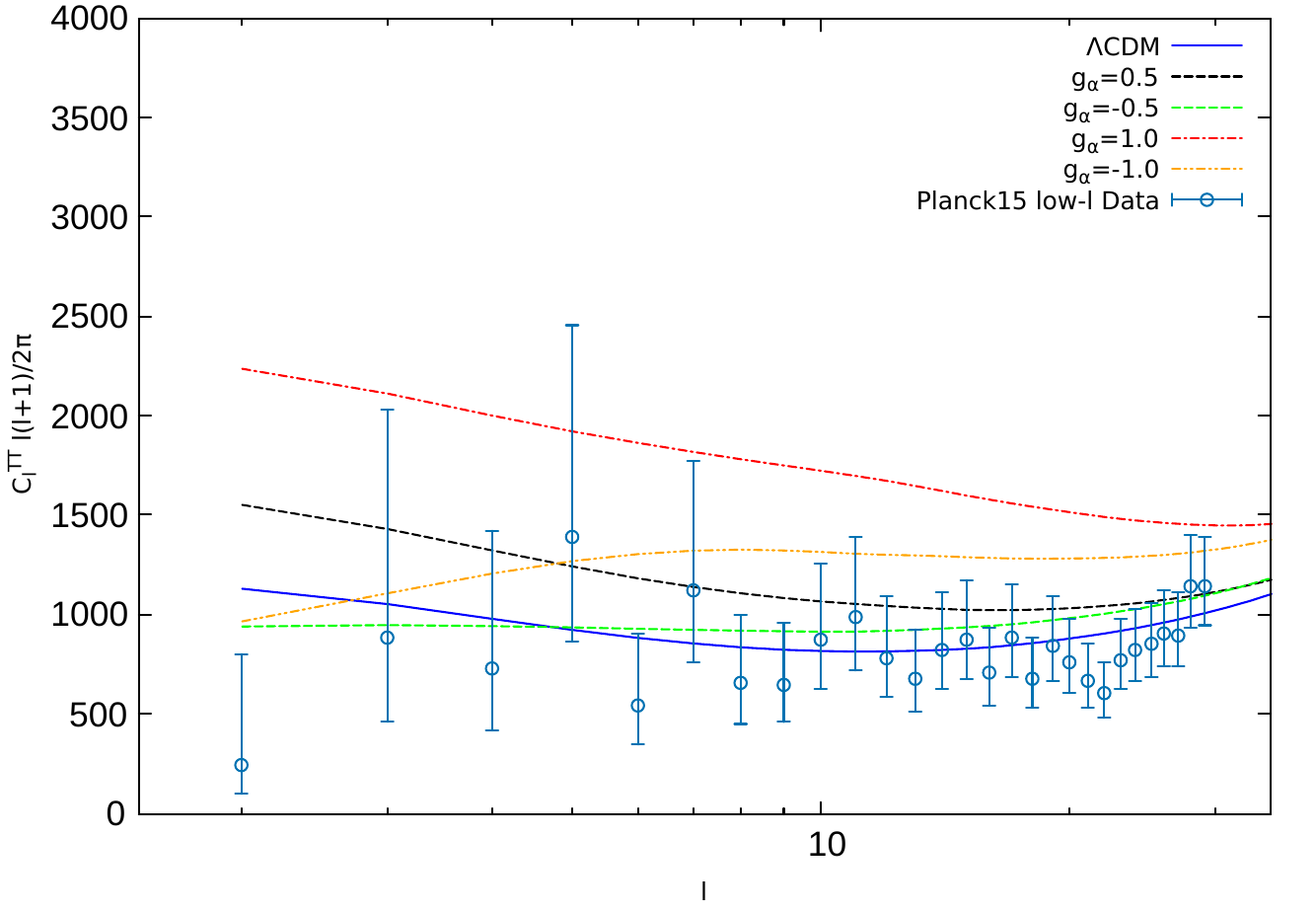}
\caption{The theoretically predicted form of the CMB power spectrum for a Planck15/$\Lambda$CDM background in the context of a varying $\mu$ cosmology described by  eq. \eqref{PKeq:geffansatz} for various values of $g_a$ (obtained using MGCAMB).}
\label{fig:mgcamb}
\end{figure}

As we can see, the low $l$ Planck data do not allow significant variations in the parameter $g_a$ and imply strong constraints on it. These constraints can be made precise using MGCOSMOMC \cite{Hojjati:2011ix,Zhao:2008bn}, the 2019 modified version \cite{Zucca:2019xhg} of the COSMOMC code \cite{Lewis:2002ah}. Allowing variation of the parameters $\left(\Omega_{0m}, \sigma_8, g_a\right)$ while fixing the rest to their Planck15/$\Lambda$CDM values we obtain the parameter contour constraints shown in Fig. \ref{fig:mgcosmomccontplc}. 

\begin{figure}[h]
\centering
\includegraphics[width = 0.8\textwidth]{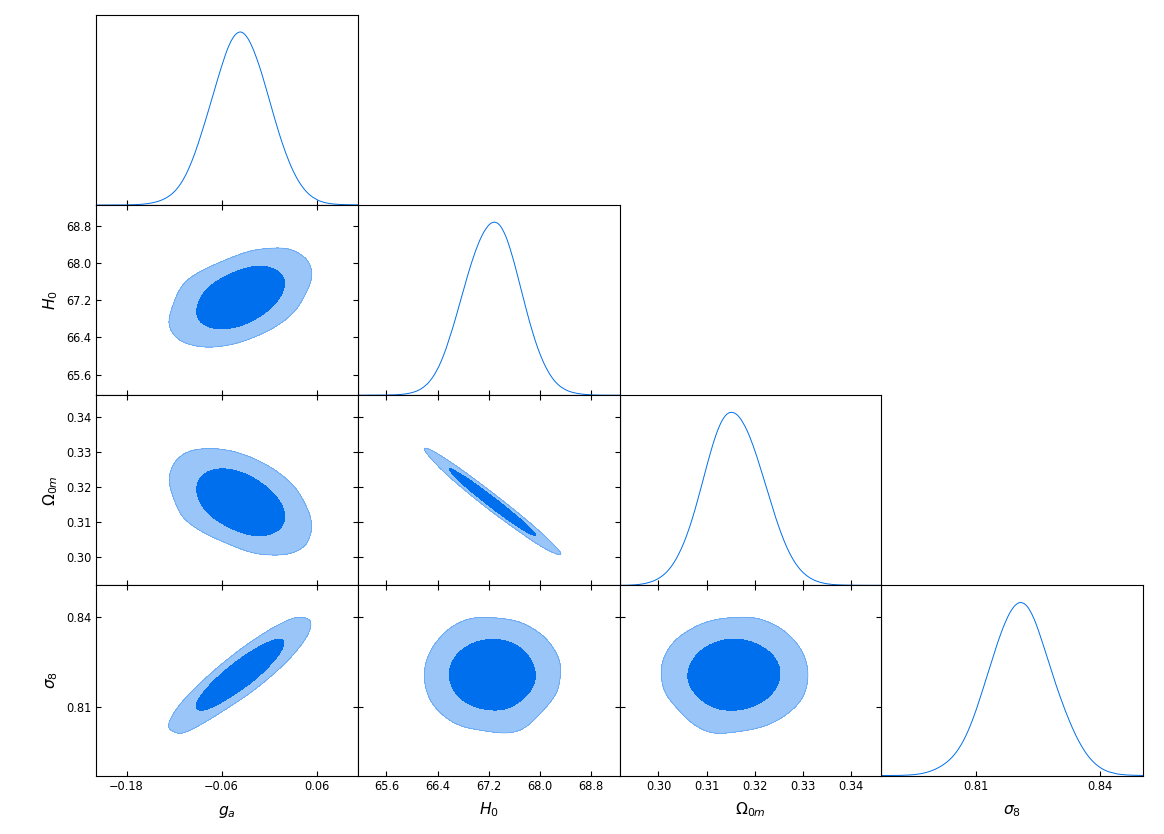}
\caption{The $1\sigma-2\sigma$ contour ranges of cosmological parameters in the context of the parametrization  \eqref{PKeq:geffansatz}, using the Planck15/$\Lambda$CDM data and setting $n=2$.}
\label{fig:mgcosmomccontplc}
\end{figure}

\noindent Clearly, even though negative values of $g_a$ are mildly favored and are consistent with the small $\mu$ variation implied by the Pantheon SnIa data of the previous section, this parameter is constrained to be larger than $-0.1$ at a $3\sigma$ level. This range is barely overalaping with the $2\sigma$ range of $g_a$ indicated by the compilation of the RSD growth data shown in Fig. \ref{fig:mtrsdcontour}. Thus, as pointed out also in previous studies \cite{Nesseris:2017vor} the low $l$ CMB spectrum constrains strongly the evolution of $\mu$ in the context of the parametrization \eqref{PKeq:geffansatz} and implies that additional parameters and/or systematic effects are required for the resolution of the weak growth tension (\eg the extension of the \lcdm $H(z)$ to $wCDM$ or the introduction of a sterile massive neutrino).

\begin{figure}[h]
\centering
\includegraphics[width = 0.7\textwidth]{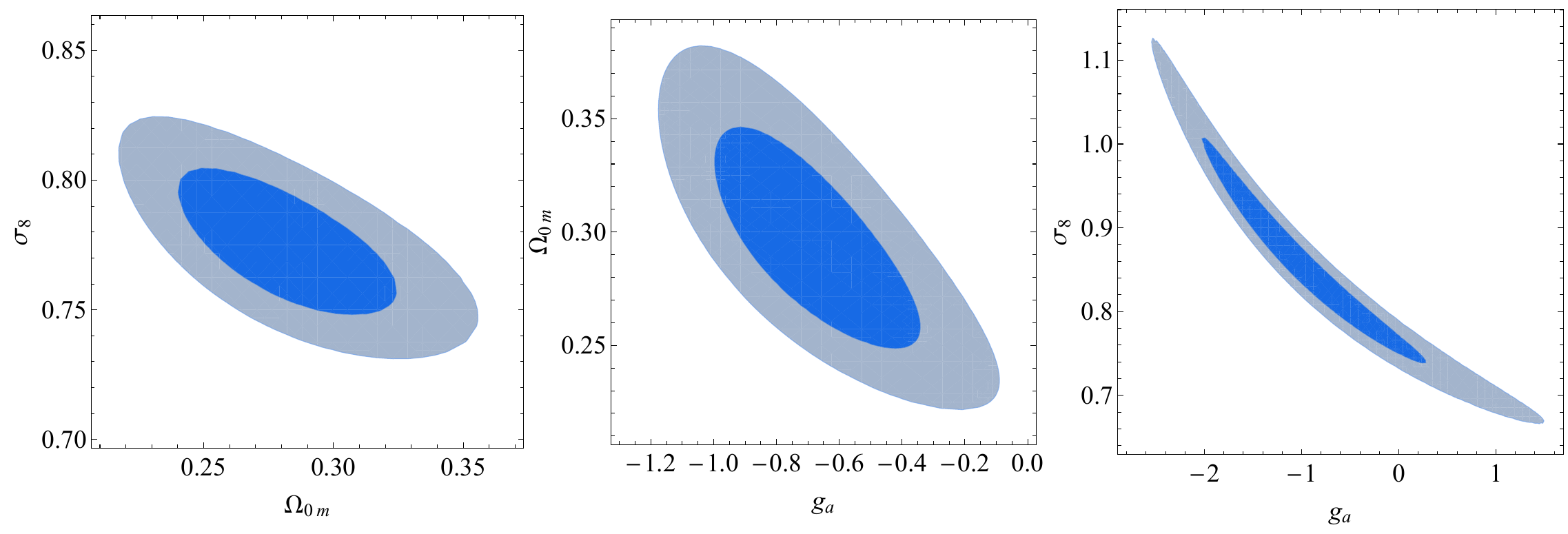}
\caption{The $1\sigma-2\sigma$ parameter constraints in the context of an evolving $\mu$ described by the parametrization  \eqref{PKeq:geffansatz} with $n=2$. The full RSD data compilation of Ref. \cite{Kazantzidis:2018rnb} was used. The third parameter in each plot was fixed to the corresponding Planck15/$\Lambda$CDM. Notice the strong indication for weaker gravity at low $z$ whose magnitude is marginally consistent the corresponding indication from CMB data (Fig. \ref{fig:mgcosmomccontplc}).}
\label{fig:mtrsdcontour}
\end{figure}

\section{Conclusions}
\label{sec:Conclusions}

Assuming a Planck/$\Lambda$CDM background expansion $H(z)$ and fixing the slip parameter $\eta$ to unity, we have investigated the constraints on a possible evolution of Newton's constant expressed through the parameter $\mu$ using three observational probes: large RSD data compilations, the Pantheon SnIa distance indicators and the TT CMB angular power spectrum from the Planck mission.  We have shown that all three probes mildly favor a Newton's constant that is weaker at low $z$ compared to GR. For RSD data this trend is at the $2-3\sigma$ level at $z<0.3$, for SnIa it is at about $2\sigma$ at $z<0.1$ and for the CMB it is at less than $1\sigma$. In the case of RSD and CMB data we have assumed a specific parametrization that respects the Solar System and nucleosynthesis constraints while reducing to GR at $z=0$ and at high $z$. The magnitude of suggested and allowed variation of $\mu$ is much smaller for the SnIa and CMB data ($1-2\%$) compared to the corresponding magnitude suggested by the RSD data (about $50\%$). This inconsistency suggests that a  variation of Newton's constant in the context of a modified gravity scenario for the parametrization and the background considered may not by itself be able to explain the weak growth tension indicated by dynamical observational probes.

The simultaneous mild indication for weaker gravity at low $z$ by  independent probes suggests the more careful investigation of the scenario of an evolving Newton's constant in the context of different $\mu$ and $\eta$ parametrizations, different $H(z)$ backgrounds and further dynamical observational probes. Such probes may include dynamical probes such as updated RSD data, WL and CC, as well as geometrical probes including CMB spectrum peaks, BAO and updated SnIa datasets.

The difficulty of viable modified gravity theories ($f(R)$ and scalar-tensor, Horndeski and beyond Hornseski) to provide a weaker gravity at low redshifts is an interesting point that may be used as a powerful discriminator among modified gravity theories.

\section*{Acknowledgements}
We thank Levon Pogosian, Alex Zucca, Savvas Nesseris and George Alestas for their help with the MGCAMB and MGCOSMOMC packages. All the runs were performed in the Hydra Cluster of the Institute of Theoretical Physics (IFT) in Madrid. This review has benefited from COST Action CA15117 (CANTATA), supported by COST (European Cooperation in Science and Technology). This research is co-financed by Greece and the European Union (European Social Fund- ESF) through the Operational Programme ``Human Resources Development, Education and Lifelong Learning" in the context of the project ``Strengthening Human Resources Research Potential via Doctorate Research" (MIS-5000432), implemented by the State Scholarships Foundation (IKY).

\raggedleft
\bibliography{Bibliography}

\end{document}